\documentclass[reprint,pra,aps,notitlepage,nofootinbib]{revtex4-2}
\usepackage{amsmath}
\usepackage{amssymb}
\usepackage{bm,braket}
\usepackage{float} 
\usepackage{tikz} 
\usepackage{verbatim}
\usetikzlibrary{quantikz}

\usepackage{hyperref}
\usepackage{multirow}
\hypersetup{
     colorlinks   = true,
     citecolor    = blue
}
\def\be{\begin{equation}}
\def\ee{\end{equation}}
\def\bea{\begin{eqnarray}}
\def\eea{\end{eqnarray}}
\def\bes{\begin{eqnarray}}
\def\ees{\end{eqnarray}}
\def\bi{\begin{itemize}}
\def\ei{\end{itemize}} 

\usepackage{amsthm}

\theoremstyle{definition}

\newcommand{\vett}[1]{{\bf{#1}}}
\newcommand{\bts}{\vett{x}}
\newcommand{\groundstate}{\mathrm{G}}

\usepackage{tikz}
\usetikzlibrary{braids}

\begin{document}
\title{Towards Studying Superconductivity in the Fermi-Hubbard Model on Rydberg Atoms } 

\author{K\"ubra Yeter-Aydeniz}\email{kyeteraydeniz@mitre.org}\affiliation{Quantum Information Sciences, Optics, and Imaging Department, The MITRE Corporation, 7515 Colshire Drive, McLean, Virginia 22102-7539, USA}

\author{Nora Bauer \thanks{Corresponding author}}\email{nbauer@mitre.org}\affiliation{Quantum Information Sciences, Optics, and Imaging Department, The MITRE Corporation, 7515 Colshire Drive, McLean, Virginia 22102-7539, USA}


\date{\today}
\begin{abstract}
We present a method for calculating the ground state energy of the Fermi-Hubbard model leveraging Rydberg atom processors and sample-based quantum diagonalization (SQD). By exploiting the perturbative relationship between the Fermi-Hubbard and Heisenberg models, the procedure samples from the Heisenberg model as prepared on the Rydberg atom processor, and uses the samples to diagonalize the Fermi-Hubbard model for large $U$. We include anisotropy and next-nearest-neighbor interactions and discuss the relevant regime for quasi-superconductivity in the 1-dimensional Fermi-Hubbard model. Numerical and experimental results on the Aquila quantum processor are presented for ground state energy calculations as well as the chemical potential. We find that the Heisenberg model sampling in the studied regime is sufficient to converge near to the ground state for up to $56$ qubits, and we see a clear advantage of Rydberg atom sampling as opposed to random sampling even with 10$\times$ more samples for diagonalization. We also present a gate-based implementation of the gate-based SQD algorithm on IBM Quantum hardware for 56-qubit Hubbard model as a benchmark. Finally, we provide a gap analysis for studying emergent superconductivity using this method. 
\end{abstract}
\maketitle 
\twocolumngrid 
\section{Introduction} 
The Fermi Hubbard model is central to understanding phenomena surrounding high temperature superconductivity. 
Classical methods for studying the Fermi Hubbard model include dynamical mean field theory (DMFT) \cite{nomura2025strongcouplinghightrmcsuperconductivity,nomura2025strongcouplinghightrmcsuperconductivity}, density matrix renormalization group (DMRG) \cite{doi:10.1126/science.aal5304}, and quantum Monte Carlo (QMC) \cite{Song_2025}. DMFT is effective in low and high dimensional systems with local interactions, but is limited by its ability to include non-local interactions for phenomena such as d-wave superconductivity. While DMRG is very effective in 1-dimensional systems, computational cost grows rapidly in higher dimensions. QMC is effective for large, high temperature systems, but can encounter the sign problem and struggle to resolve ground state properties. 

For these reasons, studying the Fermi Hubbard model in the realm of superconductivity is a burgeoning field for quantum computation. In the future, fault tolerant estimations are well-defined to extend beyond the capability of classical simulations \cite{Kan_2025}. However, even in the noisy intermediate scale quantum (NISQ) era of quantum computation simulations of real-time dynamics using over 100 qubits are already extending beyond the size of exact classical solutions \cite{chowdhury2025quantumutilitysimulatingrealtime}. Crucial properties such as the Loschmidt amplitude \cite{PRXQuantum.5.030323} have been calculated for up to 32-qubit systems, and efficient algorithms for Green's functions calculations \cite{PhysRevA.111.062610} have been studied. For ground state calculations, realizations of ordering in different phases have been computed using up to 16 qubits using variational methods \cite{Stanisic_2022}. The Sample-Based Quantum Diagonalization (SQD) method has been applied to systems up to 36 qubits on IBM Quantum hardware to compute ground state energies and superconducting correlation function \cite{PhysRevResearch.6.033107}. 

In this work, we perform ground state calculations of the Fermi-Hubbard model using SQD by exploiting the relationship between the Fermi-Hubbard and Heisenberg models in the perturbative large-$U$ limit \cite{10.1119/1.10537,JAKUBCZYK2022347}. We prepare the Heisenberg model on the quantum hardware and use it for sampling to estimate the ground state of the Fermi-Hubbard model. The anisotropic Heisenberg model has been previously studied on Rydberg atom hardware \cite{Kim_2024}. For the Heisenberg model preparation, we use a variational quantum imaginary time evolution (VQITE) \cite{wang2023varqite} to prepare an approximation to the ground state on QuEra's Aquila Rydberg atom processor \cite{wurtz2023aquilaqueras256qubitneutralatom}. Then, using a mapping between spins and 2-fermion sites, we perform the SQD protocol to estimate the ground state energy and properties of the Fermi-Hubbard model. Similarly, as a benchmark, we run the variational quantum eigensolver (VQE)-sampled \cite{Tilly_2022} SQD method on gate-based IBM Quantum hardware for 56-qubit Hubbard model and present results for a comparison. 

We compute the ground state energy and chemical potential of the Hubbard model using this sampling-based approach for systems with up to 56 orbitals, which is, to our knowledge, the largest ground state Hubbard model calculation on quantum hardware to date. In addition to this, to the best of our knowledge, our work is the first demonstration of VQITE implementation on a cloud-accessed analog quantum hardware. We directly compare our VQITE-sampling approach to random sampling with up to 10$\times$ more shots (i.e., from 1,000 shots to 10,000 shots) and the VQITE-sampling method performs better in approaching both exact ground state and chemical potential energy. 

Our paper is organized as follows: In Section \ref{sec:hubbard}, we describe the Hubbard model and its perturbative relationship to the Heisenberg model, and establish the mapping between spins and 2-fermion sites. In Section \ref{sec:methods}, we describe the Rydberg Hamiltonian and define the Aquila VQITE and SQD methods. In Section \ref{sec:numerical}, we give numerical results for SQD on the Fermi-Hubbard model for ground state and chemical potential. In Section \ref{sec:experimental}, we provide experimental results for SQD for the ground state energy and chemical potential on the Fermi-Hubbard model sampled from the Aquila quantum processor as well as gate-based VQITE on IBM Quantum hardware. Finally, in Section \ref{sec:conclusions}, we summarize our results and give direction for future work. 

\section{Fermi-Hubbard Model}\label{sec:hubbard} 
Here we present the Fermi-Hubbard model for the nearest-neighbor and anisotropic next-nearest neighbor cases and provide the second-order perturbative expansion that realizes the Heisenberg and anisotropic $J_1-J_2$ Heisenberg models. We also analyze the parameter regimes that are relevant to superconducting correlations in one dimension.
 

\subsection{Nearest-Neighbor Terms} \label{sec:NN}
Consider the repulsive nearest-neighbor Fermi Hubbard model with a kinetic energy component parameterized by hopping term, $t_{ij}$ and on-site interaction strength, $U>0$, as follows.
\be H=\sum_{\langle i,j\rangle} \sum_\sigma \bigl( t_{ij}c_{i\sigma}^\dagger c_{j\sigma}+\text{h.c.}\bigr)+U\sum_{i}n_{i\uparrow}n_{i\downarrow}~.\ee
Here, $c^\dagger$ and $c$ are the fermionic creation and annihilation operators, $\langle i, j \rangle$ denotes pairs of nearest-neighbor lattice sites, $\sigma \in \{\uparrow, \downarrow\}$ is the spin degree of freedom, and $n_{i\uparrow}, n_{i\downarrow}$ are the spin-up and spin-down fermions on site $i$.

If we take $t_{ij}$ much smaller than $U$, i.e. $U \gg t_{ij}$, then to the second-order in $t/U$ in perturbation theory \cite{10.1119/1.10537}, we obtain the effective Hamiltonian 
\be H_{\mathrm{eff}}=P_0\left(\sum_{\langle kl\rangle}|t_{kl}|^2 (2\,\mathbf{S}_l\cdot\mathbf{S}_k-\frac{1}{2})/U\right)P_0~,\ee 
where $\mathbf{S}_i=(S_i^x,S_i^y,S_i^z)$ is the spin vector on site $i$, and $P_0$ is the projection operator onto the subspace which restricts the effective Hamiltonian to where there is only one electron on each site. 

\begin{figure}
    \centering
    \includegraphics[width=\linewidth]{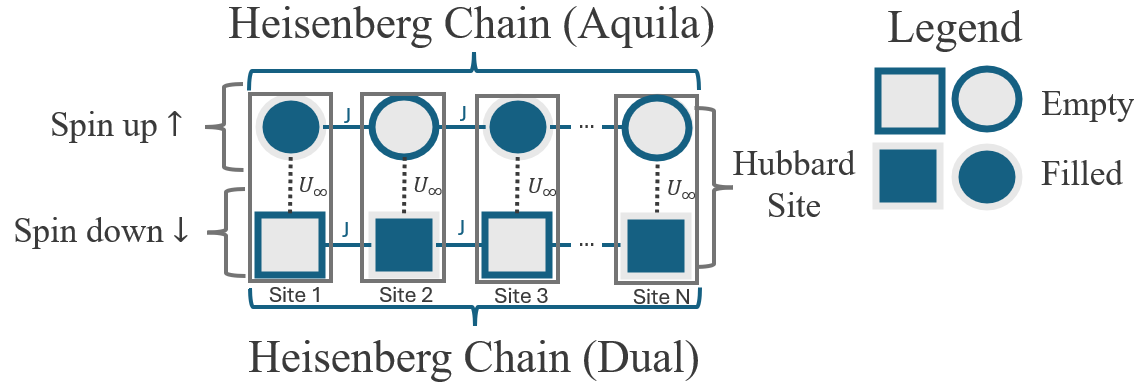}
    \caption{Schematic representation of the mapping between the two Heisenberg chains and the Hubbard model to the second order in large-$U$ ($U_\infty$) perturbation theory. Each vertical pair of spins encodes a single Hubbard site (upper: spin-up orbital, lower: spin-down orbital). The horizontal couplings $J$ implement the effective Heisenberg interactions that arise at second order in $t/U$, while a strong vertical coupling enforces the no–double-occupancy constraint.}
    \label{fig:vqite_fig}
\end{figure}

Thus we can define a mapping between two Heisenberg spin chains and the Hubbard model with respect to this second-order perturbation theory. Consider two Heisenberg spin chains side by side, as depicted in Fig.~\ref{fig:vqite_fig}. Each vertical pair represents a site of the Hubbard model, with the upper chain depicting spin up orbital and the lower chain depicting the spin down orbital. The effective Hamiltonian for each chain is the Heisenberg model, given by the horizontal $J$ interaction. There is also the constraint that there is only one electron per site, given by the vertical strong coupling, $U_\infty$, interaction. Thus, in practice, we can simulate this simplified model in this limit by simulating the upper Heisenberg chain, and, when sampling, simply inverting the measured bits for the spin down chain sites. Whenever a bitstring $x_1 x_2 \ldots x_N$ is measured on the upper chain, we construct a corresponding Hubbard configuration by assigning the complementary occupation pattern to the lower chain, i.e., $1-x_1,\,1-x_2,\ldots,1-x_N$, so that each site hosts exactly one electron.  In the example in Fig. \ref{fig:vqite_fig}, the series $101...0$ was measured for the spin up chain, so the corresponding spin down chain has $010...1$. Note that in general, this will be a large superposition of measured sequences. 
Then, we feed sampled states using this protocol into the SQD algorithm for Hubbard ground state and ground state observable estimation. 


    \label{fig:idea16}
\subsection{Next-Nearest-Neighbor Terms} \label{sec:NNN}
The nearest-neighbor, 1D Hubbard model is exactly solvable via the Bethe Ansatz. For interesting applications, especially to superconductivity, we need to move to next-nearest-neighbor (NNN) coupling. Additionally, since we would like to prepare the Heisenberg model on QuEra's Aquila quantum hardware, we need to work within the constraints of the hardware. In the testing in Section \ref{sec:methods}, we find that we can prepare the anisotropic Heisenberg (XXZ) model with Hamiltonian
\be 
H_{XXZ}=\sum_{\langle kl\rangle}J_{xy}(S_l^xS_k^x+S_l^yS_k^y)+J_{z}S_l^zS_k^z \label{XXZHam}\ee 
for anisotropy ratios $J_{xy}/J_z\leq 0.5$. Thus, we restrict our focus to the anisotropic Heisenberg model, which, as we are about to see from perturbation theory, corresponds to the spin-asymmetric Hubbard model. 

First, we can redo the perturbation theory for NNN coupled systems with spin-asymmetry. The Hamiltonian for the spin-asymmetric NNN Fermi-Hubbard system is 
\be H_{\text{Hub}}=\sum_{\langle i,j\rangle} \sum_\sigma t_{\sigma}c_{i\sigma}^\dagger c_{j\sigma}+\sum_{\langle\langle i,j\rangle\rangle} \sum_\sigma t'_{\sigma}c_{i\sigma}^\dagger c_{j\sigma}+\text{h.c.}+U\sum_{i}n_{i\uparrow}n_{j\downarrow}~.\label{eq:HubHam}\ee
Here, $t_{\sigma}$ and $t'_{\sigma}$ are the (real) hopping amplitudes for spin $\sigma\in\{\uparrow,\downarrow\}$ between nearest- and next-nearest-neighbor sites, respectively, and $U>0$ is the on-site repulsion.

If we apply the same perturbation theory for $U$ much larger than $t,t'$, we obtain the $J_1-J_2$ Heisenberg model 
\be H_{\text{eff}}=P_0\left(\sum_{kl} (2\mathcal{H}_{\text{eff}}-\frac{1}{2}(J_{z,1}+J_{z,2}))/U\right)P_0,\ee 
where
\bea \mathcal{H}_{\text{eff}}=\sum_{\langle kl\rangle}\Bigl[ J_{xy,1}(S_k^xS_l^x+S_k^yS_l^y)+J_{z,1}S_k^zS_l^z \Bigr] \nonumber\\+\sum_{\langle\langle kl\rangle\rangle}\Bigl[J_{xy,2}(S_k^xS_l^x+S_k^yS_l^y)+J_{z,2}S_k^zS_l^z \Bigr]\eea 
and the effective exchange couplings are
\begin{align}
J_{xy,1} &= \frac{4 t_{\uparrow} t_{\downarrow}}{U}, &
J_{xy,2} &= \frac{4 t'_{\uparrow} t'_{\downarrow}}{U}, \\
J_{z,1}  &= \frac{2 (t_{\uparrow}^2 + t_{\downarrow}^2)}{U}, &
J_{z,2}  &= \frac{2 \bigl((t'_{\uparrow})^2 + (t'_{\downarrow})^2\bigr)}{U}\,,
\end{align}
Thus we recover the NNN anisotropic Heisenberg model. 
Therefore, if we can prepare the NNN interacting Heisenberg model on QuEra's Aquila quantum processor, we can sample a more interesting Hubbard model perturbatively. 

\begin{figure}
    \centering
    \includegraphics[width=1\linewidth]{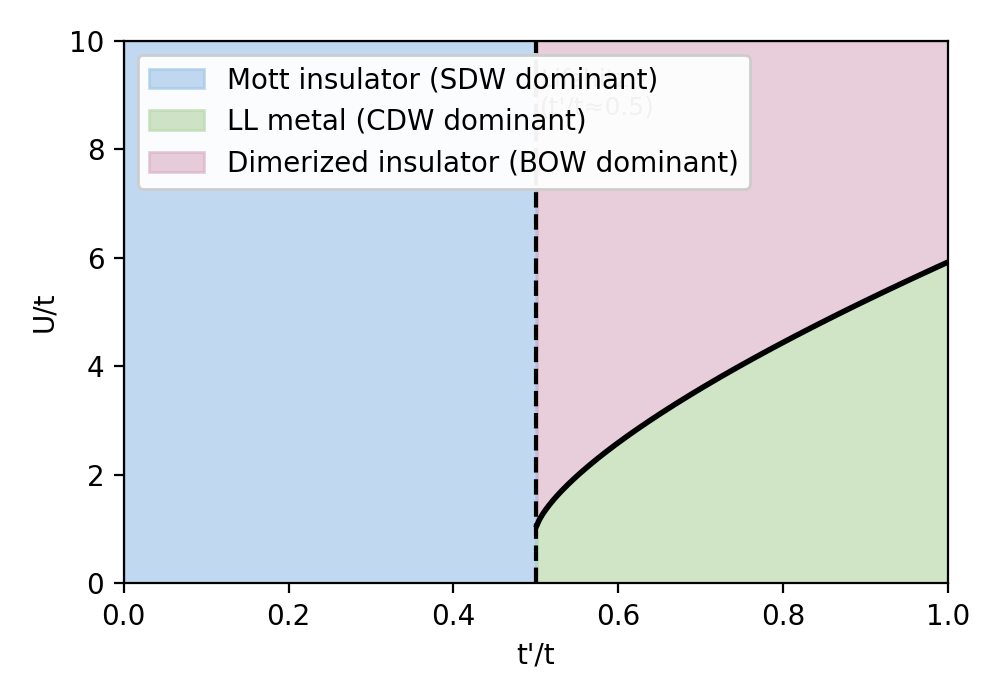}
    \includegraphics[width=1\linewidth]{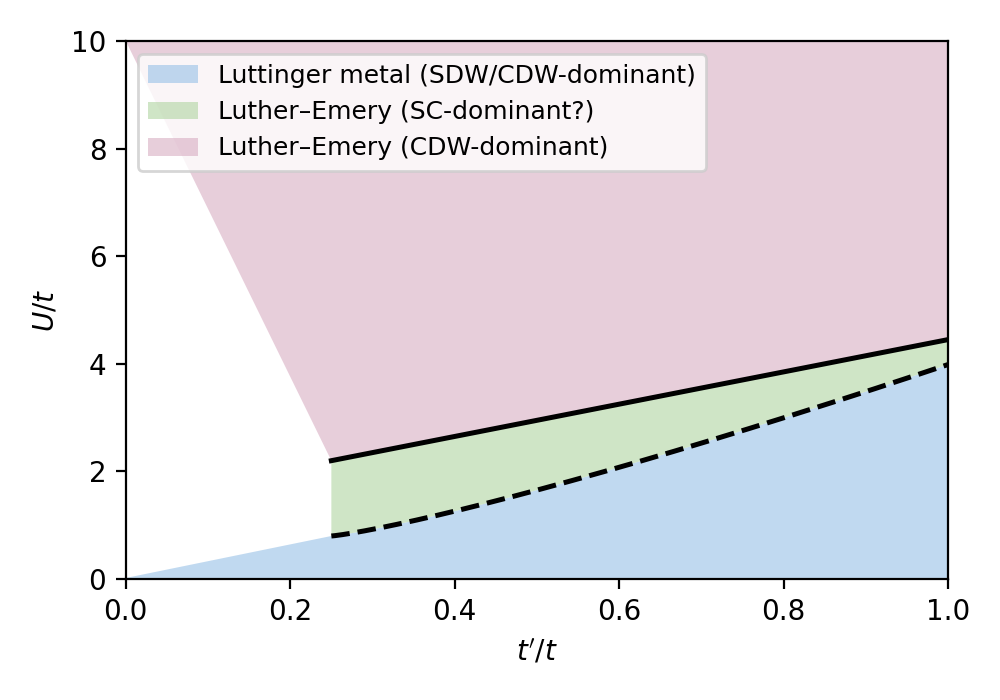}
    \caption{Schematic diagram of dominant correlations of the undoped (upper panel) and doped (lower panel) NNN neighbor 1-dimensional Fermi Hubbard model at zero temperature. 
    }    \label{fig:phasediagram}
\end{figure}

The physics gets interesting in the large $U$ limit for $t'/t\approx0.25$ \cite{PhysRevB.77.085119}. It becomes relevant for superconductivity when doped away from half-filling $5-20\%$ \cite{PhysRevB.109.085121}. 
Fig.~\ref{fig:phasediagram} provides the zero-temperature correlation regimes of this one-dimensional system as a function of $U/t$ and $t'/t$ for the undoped and doped cases. In 1D, these “phases” are not distinguished by broken symmetries but by whether the spin or charge sectors are gapped and by which correlation function has the slowest algebraic decay in the long-distance limit \cite{Giamarchi2003}. 

For the undoped model, for weak frustration $t'/t$ (blue square region in upper plot of Fig.~\ref{fig:phasediagram}) the system is described by Mott-insulating phase with quasi-long-range antiferromagnetic spin-density-wave (SDW) correlations dominant \cite{Giamarchi2003}. For larger frustration $t'/t \gtrsim 1/2$, the low $U/t$ regime (green region) becomes metallic and is well described by a Luttinger-/Luther–Emery-like liquid with charge-density wave dominant and some superconducting correlations \cite{LutherEmery1974}. As $U/t$ increases, a transition occurs into a spontaneously dimerized bond-ordered-wave (BOW) insulator (pink region). \cite{DaulNoack1998}. 

For the doped model, at weak to moderate coupling (blue region in lower plot of Fig.~\ref{fig:phasediagram}), the system is a gapless Luttinger liquid  with spin/charge density-wave correlations typically decaying more slowly than superconducting correlations. For sufficiently large $t'/t$ and/or $U/t$, the system crosses the dashed line into a Luther–Emery liquid \cite{LutherEmery1974} where the spin sector becomes gapped while the charge sector remains gapless. Depending on the doping and charge exponent $K_\rho$, superconducting correlations are the slowest-decaying (green region in Fig.~\ref{fig:phasediagram}) or charge-density wave correlations dominate (pink region in Fig.~\ref{fig:phasediagram}) \cite{DaulNoack1998}. 

It is important to note that the studied location in Hubbard parameter space is not superconducting itself, which is due to limitations from our choice of quantum hardware. The isotropic Heisenberg model ($J_{xy}=J_z$), which is related to the spin-symmetric Hubbard model ($t_\uparrow=t_\downarrow$) with phase diagram in Fig. \ref{fig:phasediagram} is not native to the Aquila quantum processor. In order to perform sampling on the Aquila processor, we consider the anisotropic Heisenberg model ($J_{xy}\neq J_z$) and the spin-asymmetric Hubbard model ($t_\uparrow\neq t_\downarrow$) as we are able to obtain this ground state with reasonable fidelity up to $J_{xy}/J_z=0.5$. Therefore, improvements in quantum hardware will enable us to move closer to the regime of superconductivity. 

\section{Methods}\label{sec:methods} 
Here we present the method for performing VQITE on the Aquila quantum processor as well as a summary of the SQD method. 
\subsection{Rydberg Hamiltonian on Aquila}
QuEra's Aquila processor works with programmable arrays of up to 256 Rubidium-87 atoms (Rb-87), trapped in vacuum by tightly focused laser beams. On Aquila processor, neutral atoms are trapped in a programmable tweezer array and driven between a ground state $|g\rangle$ and a Rydberg state $|r\rangle$ by a global laser field. In the rotating frame and under the rotating-wave approximation, the system is well described by an effective Rydberg Hamiltonian
\bea
H_{\mathrm{Ryd}}(t)
&=& \sum_{j=1}^N \Omega_j(t)\bigl(e^{i\phi_j(t)}\,|g_j\rangle\langle r_j|
      + e^{-i\phi_j(t)}\,|r_j\rangle\langle g_j|\bigr) \nonumber\\
&&\quad - \sum_{j=1}^N \Delta_j(t)\, n_j
      + \sum_{j<k} V_{jk}\, n_j n_k\,.
\eea
where $\Omega_j(t)$ is the (time-dependent) Rabi frequency of atom $j$, $\phi_j(t)$ a controllable phase, $\Delta_j(t)$ is the laser detuning frequency, $n_j = |r_j\rangle\langle r_j|$ is the Rydberg number operator on site $j$, and $V_{jk}$ the van der Waals interaction potential between Rydberg excitations on sites $j$ and $k$ \cite{wurtz2023aquilaqueras256qubitneutralatom} defined as
\be
V_{jk}=\frac{C_6}{|\vec{r}_j-\vec{r}_k|^6}~,
\ee
where $C_6=862690 \times 2 \pi$ MHz $\mu$m$^6$.

Identifying the two-level system $\{|g\rangle,|r\rangle\}$ with a spin-$\tfrac12$ degree of freedom via $|g\rangle\leftrightarrow|\!\downarrow\rangle$, $|r\rangle\leftrightarrow|\!\uparrow\rangle$, we can rewrite
\bea
H_{\mathrm{Ryd}}(t)
&=& \sum_i \Bigl[\Omega(t)\bigl(\cos\phi\,\sigma_i^x + \sin\phi\,\sigma_i^y\bigr)
- \tfrac12\Delta(t)\,(\mathbb{I}+\sigma_i^z)\Bigr] \nonumber \\
&&+ \sum_{i<j} V_{ij}\,\tfrac14(\mathbb{I}+\sigma_i^z)(\mathbb{I}+\sigma_j^z)\,,
\eea
which realizes an Ising-type spin model with tunable transverse and longitudinal fields. In this work, we use piecewise-smooth control schedules $\Omega(t)$, $\Delta(t)$, and $\phi$ as the variational parameters in the VQITE algorithm (Sec.~\ref{sec:methods}), and we arrange the atoms in a one-dimensional chain so that $V_{ij}$ decays approximately as a van der Waals interaction with distance. This native Rydberg Hamiltonian is then used to approximately prepare ground states of the target anisotropic Heisenberg model, which we subsequently sample and feed into the SQD procedure.

\subsection{Variational Quantum Imaginary Time Evolution (VQITE) Algorithm} \label{sec:VQITE}
In this section, we provide a brief overview of the variational quantum imaginary time evolution (VQITE) algorithm that we utilize on simulated Aquila quantum processor for calculating the ground state energy of the models that we are studying. In quantum imaginary time evolution (QITE) algorithm the quantum state, $|\phi_0\rangle$ evolves with imaginary time evolution 
\be
|\phi(\tau)\rangle = \frac{e^{-\tau H}|\phi_0\rangle}{\|e^{-\tau H}|\phi_0\rangle\|}\,,
\ee
where $H$ is the Hamiltonian and $\|\cdot\|$ denotes the Hilbert-space norm. This quantum state approaches to the ground state of the Hamiltonian as $\tau \to \infty$, as long as $|\phi_0\rangle$ has non-zero overlap with the ground state. Implementation of QITE algorithm on quantum hardware is challenging due to requiring non-unitary operations. Motta et al. proposed a QITE algorithm \cite{motta2020determining} that approximates non-unitary operations through a sequence of unitary operations so that they can be implemented on quantum hardware. In this version of QITE algorithm, the imaginary time evolution is divided into small $N$ ITE steps such that $\tau = N \Delta \tau$. However, these sequence of unitary operations might result in degradation in quantum state due to quantum hardware noise levels in current quantum hardware. Hence, VQITE \cite{mcardle2019variational} has been proposed to mitigate these issues.

In the VQITE algorithm, the imaginary time evolution is expressed in variational form. In VQITE, the target state, $|\phi(\tau)\rangle$, is approximated with a quantum ansatz that is defined as $\ket{\phi(\vec{\theta})}=V(\vec{\theta})\ket{\mathbf{0}}$, where $V(\vec{\theta})=U_N(\theta_N)...U_k(\theta_k)...U_1(\theta_1)$ is a sequence of parameterized unitary gates for $N$ ITE steps and $\theta_i$ represents the angles of the unitary gates. McLachlan's variational principle \cite{magnusson2024towards} is then used to approximate the imaginary time evolution by minimizing the distance between the QITE evolution and the approximated path in parameter space in $V(\vec{\theta})$ through
\begin{equation}
\delta \left|\left|\left(\frac{\partial}{\partial \tau}+H-E_\tau\right)|\phi(\tau)\rangle\right|\right|=0,
\label{eq:variation}
\end{equation}
where $H$ is the effective Hamiltonian, $|||\psi\rangle ||=\sqrt{\langle \psi | \psi\rangle}$ denotes the Hilbert-space norm, and $E_\tau=\langle \phi(\tau)|H|\phi(\tau)\rangle$ is the energy expectation value at imaginary time $\tau$ (for normalized $|\phi(\tau)\rangle$).

Substituting $|\phi(\vec{\theta}(\tau))\rangle$ in \eqref{eq:variation}, VQITE algorithm transforms into solving an equation of the form 
\be \sum_j A_{ij}\dot{\theta}_j=C_i~, \label{eq:linear_equation}\ee 
where 
\bea A_{ij}=\Re\left(\frac{\partial \bra{\phi(\tau)}}{\partial \theta_i}\frac{\partial \ket{\phi(\tau)}}{\partial \theta_j}\right)~, \nonumber\\ 
C_{i}=\Re\left(-\sum_\alpha\lambda_\alpha\frac{\partial \bra{\phi(\tau)}}{\partial \theta_i}h_\alpha\ket{\phi(\tau)}\right) 
~.\label{eq:Ci}\eea
Here, $\Re[\cdot]$ denotes the real part of the expression, and we assume $H$ has the form $H=\sum_\alpha\lambda_\alpha h_\alpha$, where $\lambda_\alpha$ are scalar coefficients and $h_\alpha$ are Hermitian operators, i.e., Pauli string operators or local interaction terms. 
We can construct our ansatz $V(\vec{\theta})$ from one- and two-qubit unitary gates. For example, for a single qubit rotation gate, $U_i(\theta_i)=R_{\theta_i}^Y=e^{-i\theta_i\sigma_Y/2}$, we have 
\be \frac{\partial U_i(\theta_i)}{\partial \theta_i}=-i/2\times Y e^{-i\theta_i\sigma_Y/2}~. \label{eq:gbansatz}\ee 
The success of the VQITE algorithm in convergence to the ground state strongly depends on the ansatz choice. The more expressive the VQITE ansatz is the higher probability of convergence to the ground state energy.
In this work, we realize $V(\vec{\theta})$ as an analog ansatz that is implemented via time evolution under a time-dependent Rydberg Hamiltonian with global control fields on Aquila analog quantum processor. The control schedule, i.e., our analog ansatz, we use is

\be \Omega(t)=\Omega_{max}\sin(\pi t)^2~,\ee
\be \Delta(t)=\Delta_{\text{start}}+(\Delta_{\text{end}}-\Delta_{\text{start}}) t~, \ee 
and 
\be \phi(t)=\phi \ee
with set of hyperparameters $(\Omega_{max},\Delta_{\text{start}},\Delta_{\text{end}},\phi,t_{max})$. Here, $\Omega_{max}$ is the maximum Rabi frequency, $\Delta_{\text{start}}$ and $\Delta_{\text{end}}$ are the initial and final detunings, $\phi$ is a global phase and $t_{max}$ is the total evolution time. We choose the initial values of these hyperparameters as follows. $\Omega_{max}= 10$ MHz, $\Delta_{\text{start}}=-12$ MHz, $\Delta_{\text{end}}=12$ MHz, $\phi=0$, and $t_{max}=0.5 \mu$s.

\begin{figure}
    \centering
    \includegraphics[width=0.95\linewidth]{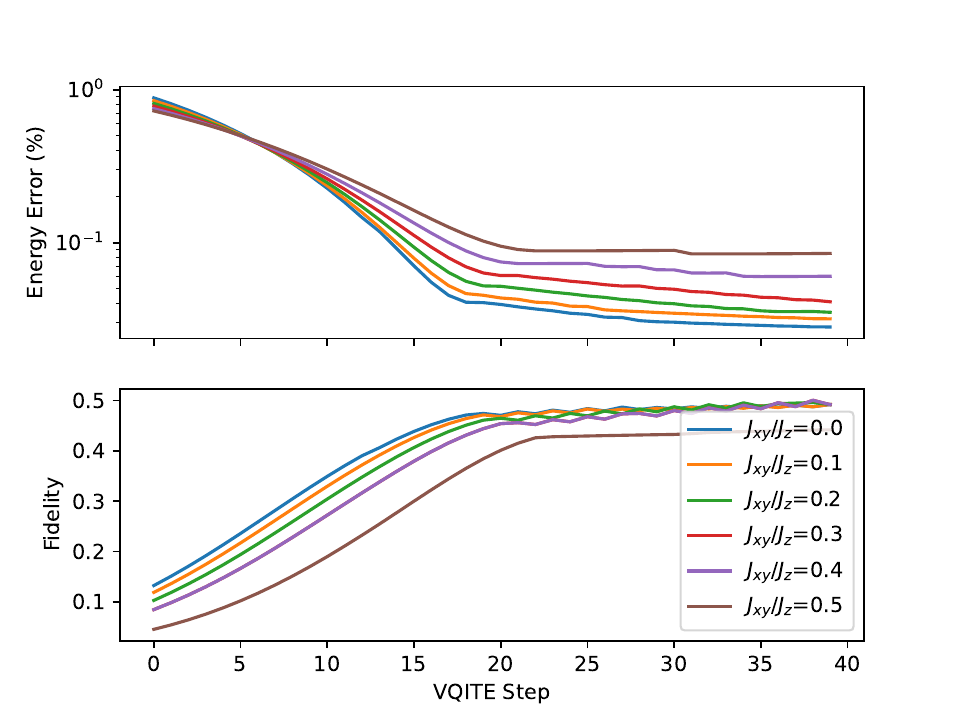}
    \caption{Energy error between ground state energy and VQITE calculated energy and fidelity between the ground state and the quantum state at given VQITE step for the $N=4$-qubit XXZ model as a function of off-diagonal coupling $J_{xy}/J_z=0.0, 0.1, 0.2, 0.3, 0.4, 0.5$, respectively.  }
    \label{fig:xxz}
\end{figure}

We utilized the VQITE algorithm and simulated this algorithm on analog quantum hardware for calculating ground state energies for the models of interest. In Fig.~\ref{fig:xxz} we show results for ground state energy error from the exact value (upper panel) and the fidelity between the ground state and quantum state at a given VQITE step (lower panel) as a function of VQITE step for this ansatz with a 4-qubit anisotropic Heisenberg (nearest-neighbor $XXZ$) model example with Hamiltonian in \eqref{XXZHam} for off-diagonal coupling of $J_{xy}/J_z=0.0, 0.1, 0.2, 0.3, 0.4, 0.5$, respectively. From these figures, we can see that VQITE algorithm that ran up to 40 steps can achieve $\sim$ 50\% fidelity for the $L=4$ anisotropic, NN XXZ model with off-diagonal coupling up to $J_{xy}/J_z\approx0.4$. 
At $J_{xy}/J_z\approx0.5$, the fidelity begins to decrease.  This motivated us to utilize SQD method on top of VQITE algorithm to achieve better accuracy on ground state energy calculations of NN XXZ model.

\begin{figure}
    \centering
    \includegraphics[width=0.95\linewidth]{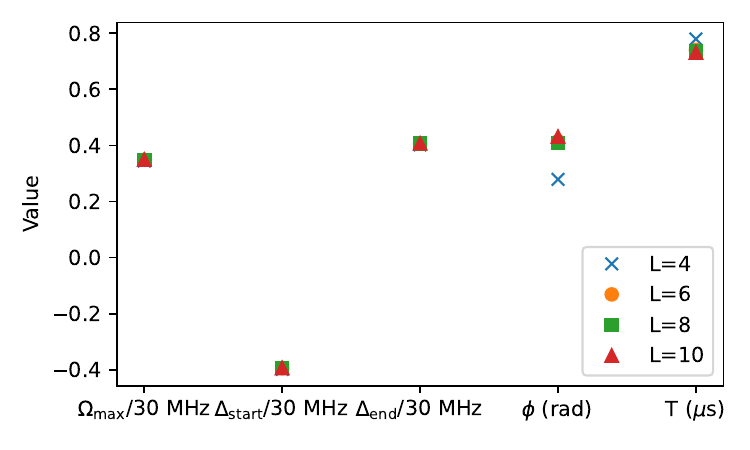}
    \caption{Convergence of the model hyperparameters for VQITE on the anisotropic NNN Heisenberg model as a function of the problem size. }
    \label{fig:hej1j2p}
\end{figure}

We implemented VQITE algorithm on Aquila simulator to calculate ground state energies of NNN XXZ model with system sizes $L=4, 6, 8, 10$, respectively. As a result of these experiments we found out that the model hyperparameters from VQITE implementation on NNN XXZ model converge as a function of the system size, as shown in Fig. \ref{fig:hej1j2p}. The converged values of the hyperparameters are $\Omega_{max} \approx 10.43$ MHz, $\Delta_{\text{start}} \approx -11.84$ MHz, $\Delta_{\text{end}}\approx 12.16$ MHz, $\phi\approx 0.38$ rad, and $t_{max}\approx0.75 \mu$s. Thus, we can fit and extrapolate parameters for larger $L$ and do not have to run VQITE again for larger $L$.

\subsection{Sample-Based Quantum Diagonalization (SQD)} 
Here we outline the sampled-based quantum diagonalization (SQD) algorithm as proposed in Ref.\ \cite{Robledo_Moreno_2025}. In this work, the algorithm was implemented on gate-based, digital quantum hardware. The authors chose to use the local unitary coupled Jastrow (LUCJ) Ansatz as their parametrized quantum circuit ansatz and optimized the parameters on classical MPS (matrix product state) software. Then the quantum circuit with optimal parameters was executed on a gate-based, digital quantum computer and the output state $|\Psi\rangle$ was measured in the computational basis $N_s$ times to obtain measurement results
\begin{equation}
    \widetilde{\mathcal{X}} = \left\{ \bts \; | \; \bts \sim \widetilde{P}_\Psi (\bts) \right\}
\end{equation}
in the form of bitstrings $\bts \in \{0, 1\}^M$ distributed according to some $\widetilde{P}_\Psi$. The bitstrings represent electronic configurations
In \cite{Robledo_Moreno_2025}, since they studied chemistry simulations, these bitstrings represent Slater determinants, i.e., occupation-number configurations in a chosen orbital basis. In our setting, we apply SQD to spin chains and to the Hubbard model in the large-$U$ mapping, hence the bitstrings measured from the analog quantum hardware correspond to the spin configurations in the computational basis of the anisotropic $J_1-J_2$ XXZ model. These spin configurations are then interpreted as the occupation patterns in the Hubbard Fock basis.

Using $K$ batches of $d$ configurations $\mathcal{S}^{(1)}\hdots , \mathcal{S}^{(K)}$ taken from the measured set $\mathcal{X}_{\textrm{R}}$, the Hamiltonian is projected and diagonalized over each $\mathcal{S}^{(k)}: k = 1, \hdots, K $. 
Each batch spans a subspace $\mathcal{S}^{(k)}$ in which the many-body Hamiltonian is projected:
\begin{equation}\label{eq:projection}
    \hat{H}_{\mathcal{S}^{(k)}} = \hat{P}_{\mathcal{S}^{(k)}} \hat{H}  \hat{P}_{\mathcal{S}^{(k)}} \textrm{, with } \hat{P}_{\mathcal{S}^{(k)}} = \sum_{\bts \in {\mathcal{S}^{(k)}}} | \bts \rangle \langle \bts | \;.
\end{equation}
The ground states and energies of $\hat{H}_{\mathcal{S}^{(k)}}$, which we label $|\psi^{(k)} \rangle$ and $E^{(k)}$, respectively, are computed using the iterative Davidson method \cite{Crouzeix_1994}.
The computational cost -- both quantum and classical -- to produce $|\psi^{(k)} \rangle$ is polynomial in $d$, the dimension of the subspace.

The ground states are then used to obtain new occupancies
\begin{equation}
\label{eq:nR_def}
    n_{p\sigma}=  \frac{1}{K}\sum_{1\leq k \leq K}  \left\langle \psi^{(k)} \right| \hat{n}_{p\sigma} \left| \psi^{(k)} \right\rangle,
\end{equation}
for each spin-orbital tuple $(p\sigma)$, averaged on the $K$ batches, where $p$ is the orbital index and $\sigma$ is the spin index. These occupancies are sent back to the configuration recovery step, and this entire self-consistent iteration is repeated until convergence, realizing an SQD of the target Hamiltonian. The initial guess for the $n_{p\sigma}$ values used for the first round of recovery comes from the raw quantum samples in the correct particle sector. 

On a noiseless signal $\mathcal{X}$, it is guaranteed to succeed efficiently if the ground state has a support $\mathcal{X}_\groundstate$ of polynomial size, and if the wavefunction $|\Psi\rangle$ prepared on the quantum processor has a support similar to that of the ground state. 

\section{Numerical Results}\label{sec:numerical} 
Here we present numerical results for energy calculations and computing the chemical potential of the Hubbard model \eqref{eq:HubHam} using the VQITE-sampled state prepared on simulated Rydberg atom analog quantum hardware. In the following calculations, we use $U=10$, $t_\uparrow=1$, $t_\downarrow=0.25$, $t_\uparrow'=0.25$, and $t_\downarrow'=0.0625$, as well as open boundary conditions. We also use the Qiskit Addon: SQD \cite{qiskit-addon-sqd} and FFSim \cite{ffsim} software. 
\subsection{Ground State Energy Calculations}
\begin{figure}
    \centering
    \includegraphics[width=0.95\linewidth]{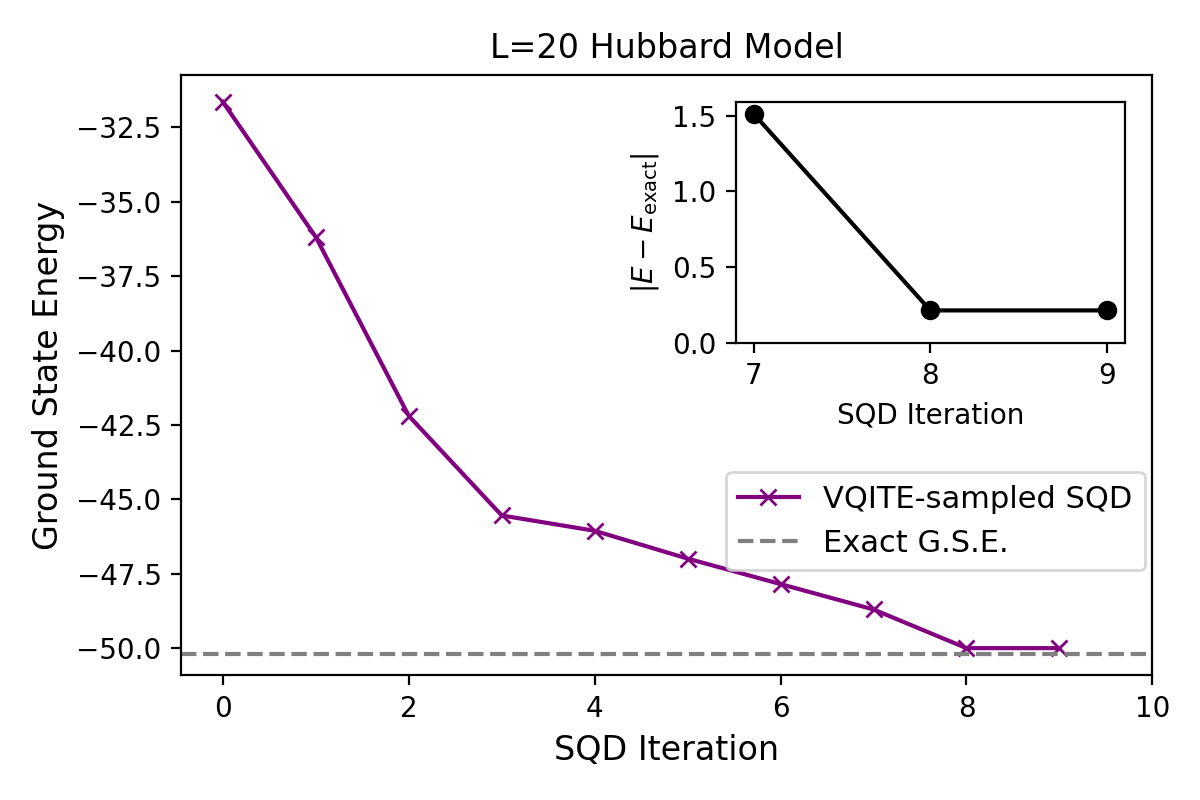}
    \includegraphics[width=0.95\linewidth]{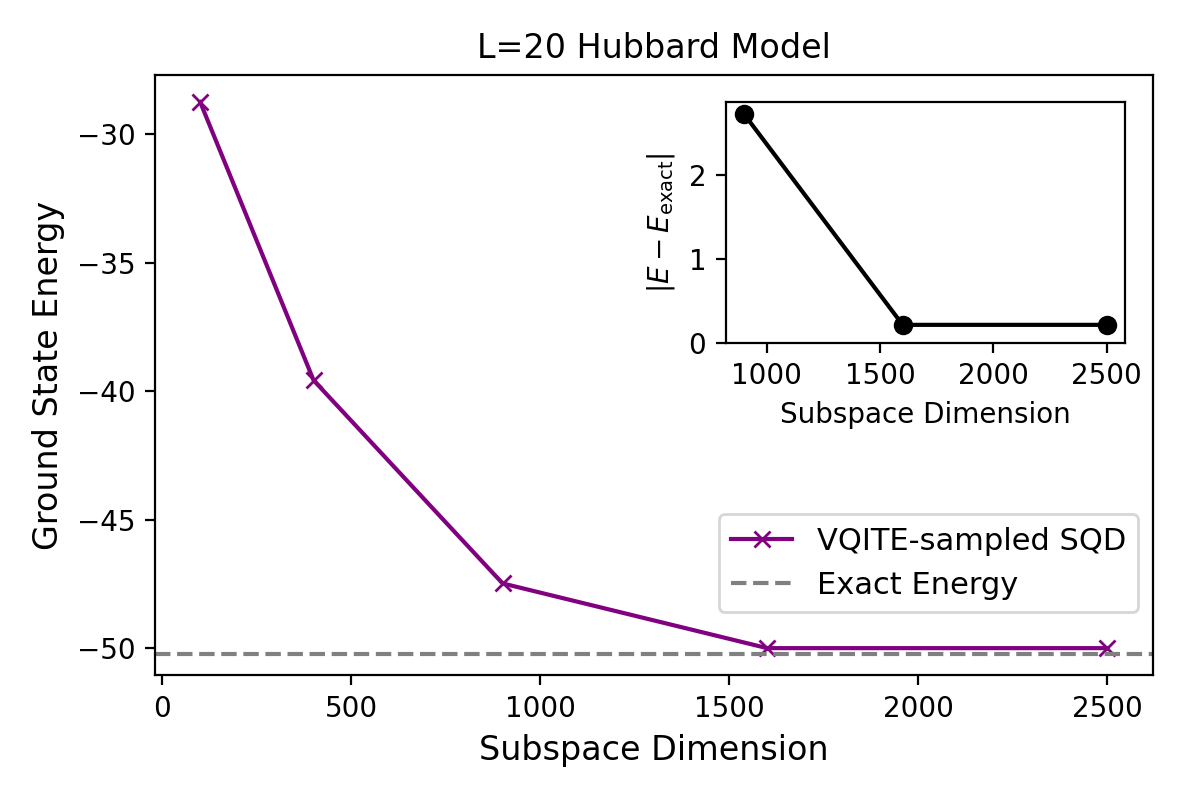}
    \caption{SQD results for the ground state energy of a 20-orbital spin-asymmetric Hubbard model at half-filling with $U=10$, $t_\uparrow=1$, $t_\downarrow=0.25$, $t_\uparrow'=0.25$, and $t_\downarrow'=0.0625$. The SQD subspaces are sampled from VQITE-prepared anisotropic NNN XXZ model ground state approximation on a simulated Aquila Rydberg processor. The upper (lower) panel demonstrates convergence to the exact ground state energy (GSE) as a function of SQD iteration (subspace dimension).} 
    \label{fig:idea}
\end{figure}

Now we can prepare an approximate ground state of the anisotropic NNN XXZ (Heisenberg) model on a simulated Aquila Rydberg processor and use samples from this state as input to SQD calculations of the ground state energy of the anisotropic Hubbard model. The performance of this scheme for the 20-orbital Hubbard model is given in Fig.~\ref{fig:idea}. The upper plot gives the performance of SQD as a function of SQD iteration using the VQITE-sampled state. The lower plot gives the energy performance as a function of the sampled subspace dimension. 

\begin{figure}
    \centering
    \includegraphics[width=0.95\linewidth]{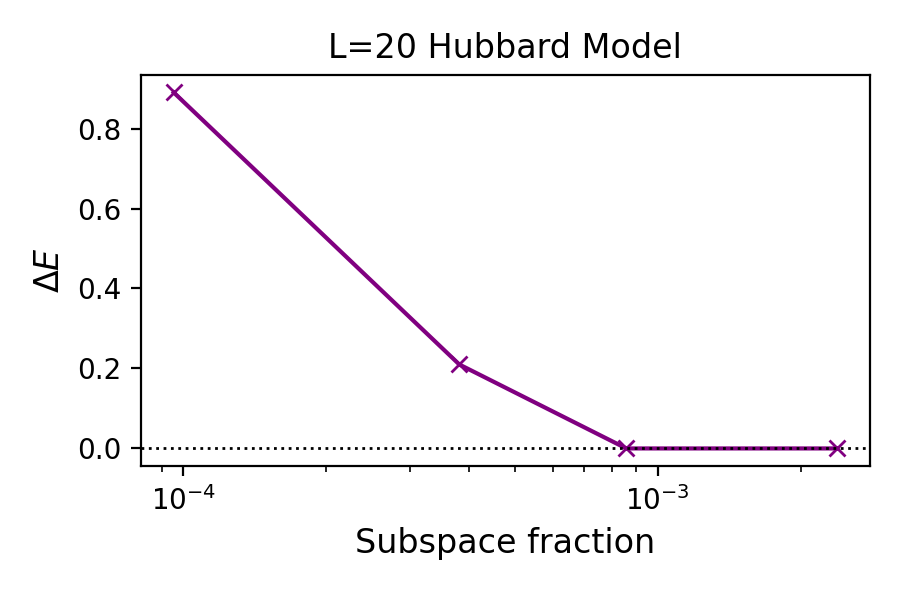}
    \caption{The difference in ground state energy error between random-sampled and VQITE-sampled SQD ($\Delta E$) as a function of the sampled subspace fraction for $L=20$ spin-asymmetric Hubbard model.
    }
    \label{fig:hubbard_advtg}
\end{figure}

Additionally, we want to compare the performance of SQD sampled from the VQITE-prepared state to SQD sampled from the uniform superposition. The result for the difference in ground state energy error between random-sampled and VQITE-sampled SQD  defined as 
\be \Delta E=\left| E_{\text{random}}-E_{\text{exact}}\right|- \left| E_{\text{VQITE}}-E_{\text{exact}}\right|
\ee
is given in Fig.~\ref{fig:hubbard_advtg} as a function of the fraction of the sampling subspace size over the Hilbert space dimension for $L=20$ spin-asymmetric Hubbard model. Here, $E_{\text{random}}$ $(E_{\text{VQITE}})$ is the ground state energy obtained using random-sampled (VQITE-sampled) SQD. For smaller fractions of the subspace, there is a clear performance benefit of sampling from the VQITE prepared state. 
\subsection{Ground State Properties} 
\begin{figure}
    \centering
    \includegraphics[width=0.95\linewidth]{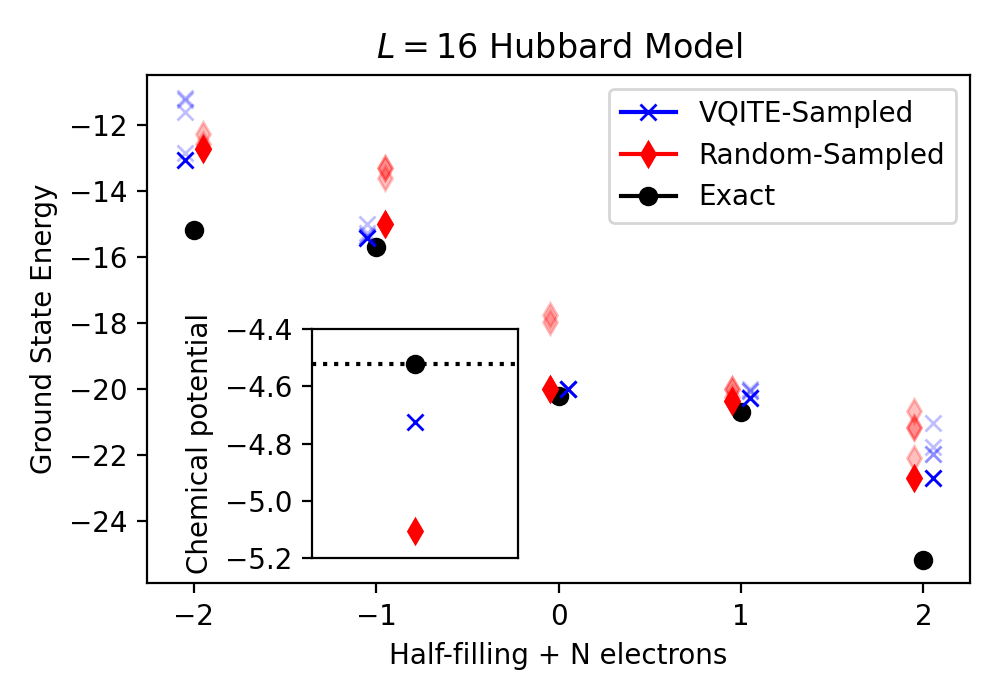}
    \caption{Ground state energies for chemical potential calculation for random- and VQITE-sampled SQD. The inset shows the chemical potential values computed from the plot's energies. The VQITE-sampled SQD results are given by the blue `$\times$'s and the random-sampled energies are given by the red diamonds. The true ground state energies, when available, are given by the black dots. Individual instances of computed energies are given by the translucent markers. }
    \label{fig:chem_potential}
\end{figure}

In addition to the ground state energy, we can also calculate energy-based observables from SQD. The chemical potential $\mu(N_{occ})$ is given by 
\be \mu(N_{occ})=E(N_{occ})-E(N_{occ}-1)~,\label{eq:chempot}\ee 
where $E(N_{\mathrm{occ}})$ denotes the ground state energy of the Hubbard model in the section with total occupation $N_{\mathrm{occ}}$ and the derivative of the chemical potential $\mu'(N_{occ})$ is given by 
\be \mu'(N_{occ})=E(N_{occ}+1)-2E(N_{occ})+E(N_{occ}-1)~.\ee 
The ground state energy values obtained from random- and VQITE-sampled SQD for various filling values for $L=16$ spin-asymmetric Hubbard model are given in Fig. \ref{fig:chem_potential}. The inset shows the chemical potential value from VQITE- and random-sampled SQD, where the VQITE-sampled value is significantly closer to the exact value. 

\section{Experimental Results}\label{sec:experimental} 
Here we present the experimental results for SQD, VQITE-sampled on QuEra's Aquila quantum processor. In the following results, we use 1000 shots for the Aquila sampling. We compare with random sampling with 1000 and 10,000 shots. In the following calculations, we use $U=10$, $t_\uparrow=1$, $t_\downarrow=0.25$, $t_\uparrow'=0.25$, and $t_\downarrow'=0.0625$, as well as open boundary conditions. The classical comparison results are computed using the density matrix renormalization group (DMRG) method using the TeNPy package \cite{tenpy2024}. 

\begin{figure}
    \centering
    \includegraphics[width=0.95\linewidth]{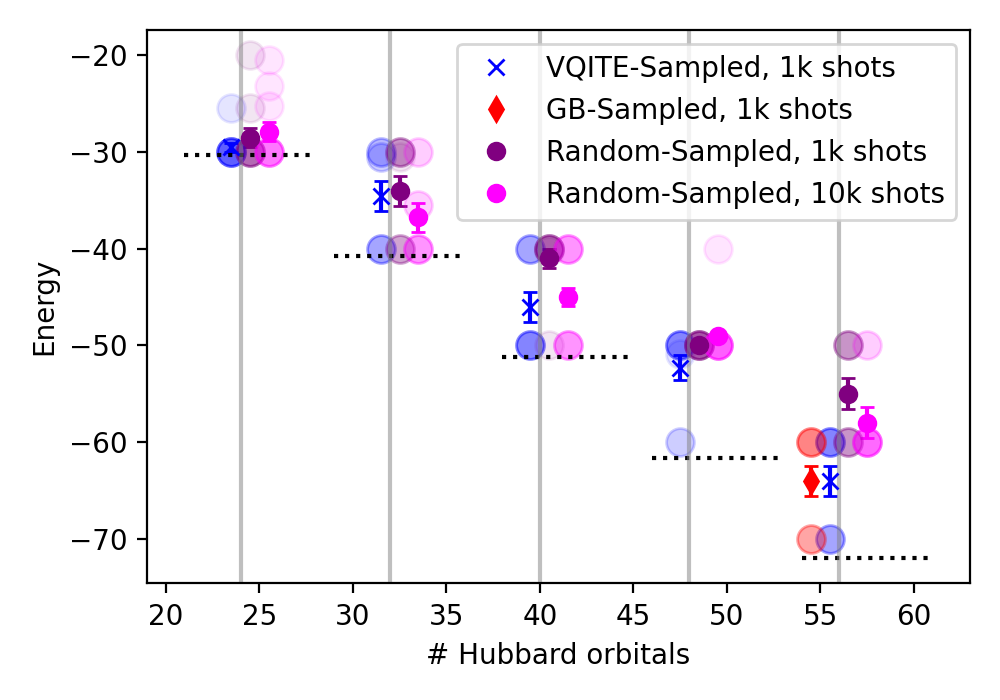}
    \caption{Performance of VQITE-sampled SQD on Aquila and random-sampled SQD for calculating ground state energies of spin-asymmetric Hubbard model with  various number of Hubbard orbitals. The VQITE-sampled SQD results are given by the blue `$\times$'s and the random-sampled energies with 1000 (10,000) shots are given by the purple (magenta) dots. The DMRG energies are given by the black dotted lines. Individual instances of computed energies are given by the large translucent dots. The ground state energy for 56-qubit case obtained using gate-based VQE-sampled SQD (denoted as GB-Sampled) on \texttt{ibm-pittsburgh} quantum computer is presented with red diamond.  }
    \label{fig:aquila_1}
\end{figure}

We prepare the approximate Heisenberg states on the cloud-accessed Aquila processor for states with 12-28 lattice sites using the extrapolated parameters, and use it to compute the ground state energies with SQD for 24-56 orbitals. We compare it to random-sampled SQD energies for 1000 and 10,000 shots. These results are given in Fig. \ref{fig:aquila_1}. 
The VQITE-sampled SQD results are given by the blue `x's and the random-sampled energies with 1000 (10,000) shots are given by the purple (magenta) dots. The ground state energies as computed with DMRG are given by the black dotted lines. Individual instances of computes energies are given by the large translucent dots.

For 24 orbitals, the energies produced from VQITE-sampled and random sampled SQD are comparable, and the gap between them increases with lattice size. By 56 lattice orbitals, there is an energy gap of almost 10 between VQITE and random-sampled energies, with VQITE finding a significantly lower ground state energy estimate. 

In addition to the analog implementation of VQITE-sampled SQD algorithm on QuEra's Aquila Rydberg processor, we present our gate-based implementation of our SQD workflow implementation on IBM Quantum superconducting hardware. These results are valuable to contribute to the benchmarking of the developed framework on various quantum hardware. The gate-based ansatz we used to conduct our experiments is given by the Qiskit Efficient SU2 Ansatz, and we used randomized initial parameters. The quantum circuit parameters, $\theta_i$, are optimized using the VQE algorithm \cite{Tilly_2022} with a maximum of 100 optimizer iterations and sampled on the 156-qubit \texttt{ibm-pittsburgh} device. We present our results for calculating ground state energy of the 56-qubit spin-asymmetric Hubbard model. We present our results in Fig. \ref{fig:aquila_1}. The results show that the ground state energy on gate-based VQE-sampled SQD method is comparable to the ground state energy obtained from the experiments run on Aquila analog quantum processor. However, both the gate-based and analog experiments outperform the random-sampled SQD. These results demonstrate that the framework we developed is a quantum hardware agnostic method and can be leveraged for studying ground state energy of Fermi-Hubbard model.

\begin{figure}
    \centering
    \includegraphics[width=\linewidth]{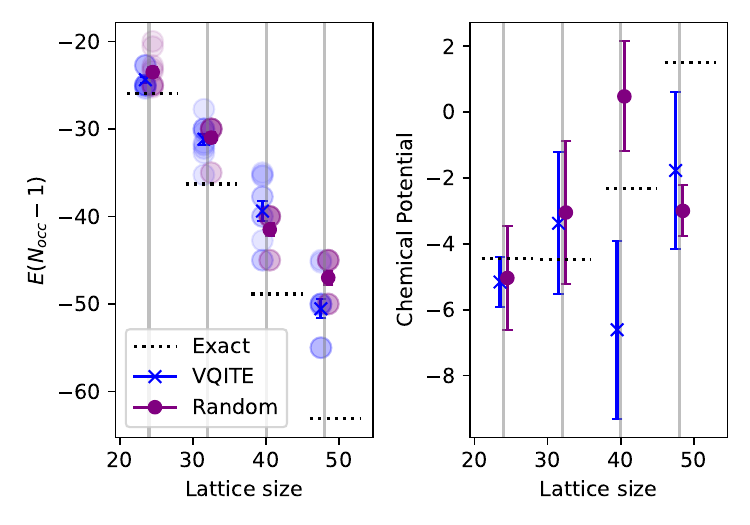}
    \caption{Results for measuring the chemical potential of the 12-site Hubbard model for VQITE (blue `$\times$') and random-sampled (purple `o') SQD compared to the exact value (dotted line). The VQITE-sampled SQD results are given by the blue `$\times$'s and the random-sampled energies with 1000 shots are given by the purple dots. The DMRG calculated ground state energies are given by the black dotted lines. Individual instances of computed energies are given by the large translucent dots.   }
    \label{fig:chem_exp}
\end{figure}

Then, we compute the chemical potential \eqref{eq:chempot} experimentally using the sampling method for 1000 shots. The results for $E(N_{occ}-1)$ and the chemical potential are given in Fig. \ref{fig:chem_exp}. For the $E(N_{occ}-1)$ calculation, the VQITE-sampled results generally have an advantage over the random-sampled results, while both struggle to get close to the DMRG value for larger systems with 1000 shots. There is not a consistent trend with the chemical potential values, which is likely due to the chemical potential being the difference of two values with error. 

\section{Conclusions}\label{sec:conclusions} 
In this work, we have used Rydberg atoms in an SQD scheme to compute ground state energy values and ground state properties of the Fermi-Hubbard model. By exploiting the large-$U$ perturbative relation between the Heisenberg and Hubbard models, we can sample on the Heisenberg model and use it to compute properties of the Hubbard model. We used VQITE to prepare a ground state approximation of the anisotropic Heisenberg model on Rydberg atoms. Then, we used the SQD protocol sampled on the VQITE-prepared state to estimate the ground state energy and chemical potential and compared directly to random-sampled results to prove the necessity of quantum sampling. On QuEra's Aquila quantum hardware, we found that the VQITE-sampled state performed better than random sampling with 10$\times$ the number of shots, and we were able to converge close to the ground state for up to 56 orbitals in the Hubbard model. 

To further extend our research, we also modified our analog VQE-sampled SQD framework to gate-based architecture and ran our experiments on cloud-accessed IBM Quantum \texttt{ibm-pittsburgh} chip. These experiments demonstrated that our framework is quantum hardware agnostic and can be extended to other quantum architecture as well. 
For further directions, it would be interesting to test further on gate-based quantum hardware because this would allow us to prepare the isotropic Heisenberg model, albeit with potentially more error. Additionally, performing SQD with time-evolved states would be a potential extension to improve results 
\cite{75pv-hbrx}. Including doping in the system would also allow the results to be pushed closer to the relevant quasi-superconducting regime, and moving to 2-dimensional systems would allow for full superconductivity to be studied. 

Another interesting line of research involves programmable Fermi-Hubbard simulators \cite{3nx4-bnyy} as an alternative to using neutral atom hardware. Using SQD or SQDOpt \cite{https://doi.org/10.1002/qute.202500248} could prove extremely powerful in pushing the boundaries of Fermi-Hubbard calculations. Additionally, quantum dots offer a promising avenue for Fermi Hubbard simulations \cite{donnellyLargescaleAnalogueQuantum2026}.

\section{Acknowledgements} 
©2026 The MITRE Corporation. ALL RIGHTS RESERVED. Approved for public release. Distribution unlimited PR 26-0348. 

This material is based upon work supported by the U.S. Department of Energy, Office of Science, Office of Advanced Scientific Computing Research, under Award Number DE-SC0024451.
The Aquila processor was accessed through Amazon Braket, and we acknowledge support from the AWS Cloud Credit for Research program. 

This research used resources of the Oak Ridge Leadership Computing Facility, which is a DOE Office of Science User Facility supported under Contract DE-AC05-00OR22725. 
\bibliography{bibliography}

\end{document}